\documentclass[lettersize,journal]{IEEEtran}
\usepackage{amsmath,amssymb,amsfonts}
\usepackage{algorithmic}
\usepackage{algorithm}
\usepackage{array}
\usepackage[caption=false,font=normalsize,labelfont=sf,textfont=sf]{subfig}
\usepackage{textcomp}
\usepackage{stfloats}
\usepackage{url}
\usepackage{verbatim}
\usepackage{graphicx}
\usepackage{cite}
\usepackage{booktabs}

\usepackage{xurl}                 
\usepackage[hidelinks]{hyperref}  
\begin{document}

\title{Grid‑Forming AC/DC Converters for Meshed DC Networks via Dual‑Measurement Frequency Droop}

\author{Hisham Mahmood,~\IEEEmembership{Senior Member,~IEEE,}
	\and
	and Marcelo A. Elizondo,~\IEEEmembership{Senior Member,~IEEE}
	\thanks{The Pacific Northwest National Laboratory 	(PNNL) is operated by Battelle for the U.S. Department of Energy under Contract DE-AC05-76RL01830. This work was supported by the Energy System Co-Design with Multiple Objectives and Power Electronics (ECOMP) Initiative at PNNL. (Corresponding author: Hisham Mahmood.)\\ Hisham Mahmood, and Marcelo A. Elizondo are with PNNL, Richland, WA 99352 USA (e-mail: hisham.mahmood@pnnl.gov; Marcelo.Elizondo@pnnl.gov).}
}

\maketitle

\begin{abstract}
	This paper proposes and analyzes a dual‑measurement frequency‑droop (DMFD) controller for meshed MVDC networks. Using only two local DC measurements, bus‑voltage deviation and a filtered DC‑port current or power deviation, the controller enables grid‑forming operation at both AC and DC ports across the mesh.
	Although the DMFD regulates neither the DC voltage nor the DC current directly and acts only on the terminal frequency, it is shown that AC-side synchronization closes an effective DC droop at every synchronized terminal, so that the steady operating point of the mesh exists and is unique. Constant frequency deviations at a terminal and single-terminal islanding enter the same DC equations as equivalent current offsets, with the island frequency fixed by the island power balance, allowing node voltages and branch currents to be obtained from one linear system.
	The steady-state model yields design-use envelopes to map the droop ratios and the net DC-injection levels into bounds on DC-voltage deviation and terminal currents.
	A small‑signal dynamic model is developed and used to generate eigenvalue loci and damping‑ratio charts that guide controller‑gain selection. Averaged EMT simulations on a four-terminal ring corroborate the analysis: in regimes where PLL-based grid-following control is known to lose damping, the DMFD controller maintains stable operation and follows the predicted steady-state characteristics.

\end{abstract}


\section{Introduction}
\IEEEPARstart{M}{ulti‑terminal} DC (MTDC) networks are emerging as enabling infrastructure for converter‑rich power systems, where short‑circuit strength is limited and large, dynamic loads concentrate behind weak interties.  
In such environments, the conventional grid‑following (GFL) paradigm, exhibits well‑documented sensitivity at low short‑circuit ratio (SCR), degraded stability margins during severe voltage depressions, and limited autonomous support during disturbances \cite{nerc_2021_gfm_whitepaper,rosso_ojia_2021_gfm_review,bahrani_pesmag_2024_gfm_landscape,wang_ojia_2020_sync,taul_tpel_2019_sync_assessment}. Recent NERC guidance on emerging large loads further documents converter‑driven instability risks of PLL‑based GFL controls at low SCR (e.g., \(\mathrm{SCR}\!\approx\!2\)), with field incidents underscoring the vulnerability of large power‑electronic loads connected to weak networks \cite{nerc_2025_large_loads}. Research studies using impedance‑based methods provide a complementary mechanism‑level explanation of poor damping and loss of synchronism in such conditions \cite{Sun2011_Impedance_TPEL}. Parallel efforts on multi‑terminal and multi‑vendor HVDC explicitly codify terminal‑level GFM functions, including frequency formation/participation, active‑power control, and fault behavior, thereby setting expectations for future MTDC deployments \cite{interopera_2024_d22,entsoe_2019_hvdc_ops,entsoe_2021_mfs_hvdc}.

On the DC side, MTDC practice commonly employs decentralized DC‑voltage/power droop to provide average DC‑bus regulation and sharing. These structures are effective for steady‑state power balancing within the DC grid, yet they do not address the lack of local AC‑side formation and therefore cannot by themselves remedy weak‑grid limitations at the AC interfaces. Small‑signal analyses of MTDC droop and its interaction with transfer limits and modes are well established \cite{PrietoAraujo2011_MTDC_Droop_TPWRD,Thams2017_DroopStructures_TPWRD}; coordination of droop with higher‑level objectives has also been examined \cite{Rouzbehi2014_OPF_Droop_JESTPE}. A unified posture that exhibits grid‑forming behavior on \emph{both} domains is therefore desirable: (i) AC‑side frequency formation/participation, and (ii) cooperative DC‑bus regulation and sharing across the mesh.

A dual-measurement frequency droop (DMFD) that moves toward this posture was recently proposed for solid-state transformers (SSTs) feeding interconnected microgrids \cite{Mahmood_SST_2026}, where the grid-side frequency reference is synthesized from the local DC-link voltage deviation and the measured grid-side power deviation, providing grid-forming capability at the AC terminal without a PLL or mode switching. The analysis in \cite{Mahmood_SST_2026} is, however, specific to the SST configuration: each DC link is a terminal-local energy buffer isolated from all others, so its steady-state balance is set by its own feeder and the DC-voltage relation decouples unit by unit; units coordinate only through a common AC bus, so the synchronous frequency is a shared signal and any deviation in it is common-mode; and islanding is a system-wide event in which all units lose the upstream anchor together. None of these structural and operational properties survive in a meshed MVDC network. The DC bus becomes a node of a resistive network whose voltage depends on every injection and every branch resistance, so existence and uniqueness of the operating point under distributed droop must be established rather than assumed. Terminals may be tied to different AC systems, so a steady frequency bias at one AC port is a differential disturbance transmitted through the DC mesh to all other terminals, and energy balance becomes a network-level rather than a local condition. A single terminal may island while the rest remain anchored, raising the question of whether the others absorb it within DC-voltage and current limits. In addition, injections exceeding a terminal's capacity can drive the droop into infeasible regions, and DC capacitors and line impedances introduce inter-terminal modes with no counterpart among isolated DC links.

This paper therefore studies the DMFD controller for the terminals of a meshed MVDC network, using only two local DC measurements at each terminal: DC-bus-voltage deviation and a filtered DC-port current (or power) deviation.

In this work, the DMFD frequency command is integrated directly to generate the converter voltage angle, so that no PLL is required in the primary control loop, and the resulting voltage reference is applied through a cascaded voltage controller with current limiting, as shown in Fig.~\ref{fig:ctrl}. Other grid-forming realizations that expose a frequency-to-active-power path, such as power-synchronization control, virtual synchronous machines, and virtual-oscillator control \cite{zhang_tpwrs_2010_psc,Zhong2011_Synchronverter_TIE,dhople_allerton_2013_voc,Johnson2016_VOC_TPEL}, can host the same command; the analysis and simulations in this paper are carried out for the direct angle-integration realization.
In the small‑signal model this AC‑side path is represented by the synchronizing gain \(K_{P\delta}\) that maps \(\Delta\omega-\Delta\omega_g\) to \(\dot{\Delta P}\); robustness under very‑weak interconnections has been documented for PSC implementations \cite{Zhang2011_PSC_Weak_TPWRS,wu_tie_2019_psc_stability}.

The formulation keeps the DC network in SI units on the standard
node--branch representation~\cite{cigre_tb604_2014} and performs
controller tuning in per unit, with an explicit mapping from per-unit
gains to an effective SI droop so that terminals of different ratings
share one physical network model.

\textit{Contributions:} The paper addresses whether frequency-droop
grid-forming control can be adopted at every terminal of a meshed DC
network, and at what cost, with the following results.
(i) It is shown that dual-sided grid forming does not leave the DC link
floating: because AC synchronization alone closes an effective DC droop
under DMFD, the operating point of a connected mesh exists and is unique
without any terminal regulating DC voltage or current directly. This
removes the structural objection to AC-side GFM in MTDC networks and
shows that the DC-side models used in current MTDC practice remain
valid without modification.
(ii) The price of placing frequency in the droop law is quantified. A
steady AC-frequency bias at any terminal, and the islanding of any
terminal, become DC-network quantities that redistribute power across
the mesh; the paper shows that both enter the same linear nodal
equations as equivalent current offsets, with the island frequency
fixed by local power balance, so their effect on every DC voltage and
branch current is computable in closed form rather than observed case by
case.
(iii) From the same equations, design-use envelopes and safe-band
contours relate the droop ratios and the net DC-injection levels to
DC-link bands and terminal-current limits, so that gains can be selected
and audited against operating limits before any time-domain study.
(iv) A small-signal model of the \(N\)-terminal network with DMFD
shows that under voltage-only droop the system is marginally stable,
with an undamped oscillatory mode, and that the current channel of the
DMFD damps this mode. The second measurement is therefore a stability requirement for grid-forming terminals on a mesh, not merely a sharing refinement, and the resulting damping maps over \((m_{dc},m_i)\) give the admissible gain region.
(v) Simulations on a four-terminal ring under uneven loading, islanding
of a terminal, and a three-phase fault on a weak AC tie
(\(\mathrm{SCR}\approx2\)) show that the AC-side benefits of grid
forming, including stable islanding and fault ride-through where a GFL
baseline with the same DC-droop gains loses stability, are obtained
without giving up the steady-state DC-droop behavior of conventional
MTDC control.

Section~II introduces the current‑based DMFD with terminal‑base per‑unit metering. 
Section~III develops the steady‑state model, presents operating cases for frequency deviation and islanding, and summarizes the design‑use envelopes. 
Section~IV formulates the dynamic model for an \(N\)‑terminal network; 
Section~V reports averaged EMT simulation results on a four‑terminal ring.
Section~VI concludes the study.

\section{Proposed Controller}
\begin{figure*}[t]
	\centering
	\includegraphics[width=6.0 in]{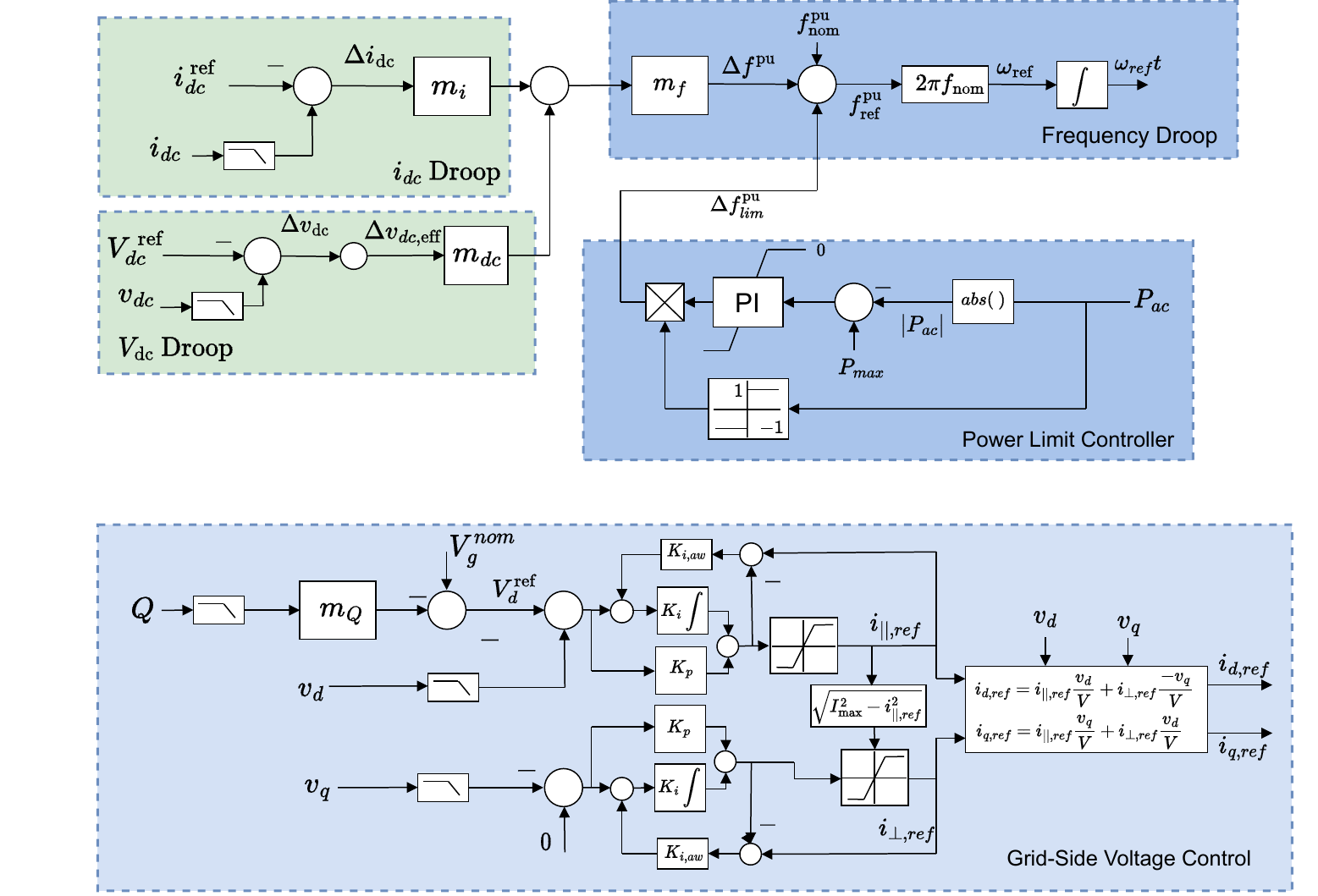}
	\caption{Block diagram illustrating the components of the DMFD controller along with the voltage controller of representative terminal MMC.}
	\label{fig:ctrl}
\end{figure*}

The converter at each terminal implements a dual-measurement frequency droop (DMFD) relation that conforms grid-forming capabilities at both terminals by synthesizing the reference-frequency adjustment from two locally measured DC-side quantities: the terminal DC-bus voltage deviation and the DC-port current deviation relative to its scheduled reference. Currents and exogenous nodal injections are defined positive \emph{into} the DC mesh; AC active power is defined positive \emph{into} the AC network. The DC-port current deviation is
\begin{equation}
	\Delta i_{dc,k} \;\triangleq\; i_{dc,k} - i_{dc,k}^{\mathrm{ref}} ,
	\label{eq:dmfd_delta_i_def}
\end{equation}

and the quantity fed to the controller, \(\Delta i_{dc,f,k}\), is the output of a
first-order low-pass filter with time constant \(T_k\) applied to
\eqref{eq:dmfd_delta_i_def}; at steady state \(\Delta i_{dc,f,k}=\Delta i_{dc,k}\).

The \emph{Current-based} DMFD implemented in this paper is shown in Fig.~\ref{fig:ctrl} along with the voltage controller of representative terminal MMC. With per-unit DMFD coefficients for voltage and current, the frequency-deviation command is

\begin{equation}
	\Delta \omega \;=\; m_{f}\Big( m_{dc}\,\Delta v_{dc} \;+\; m_{i}\,\Delta i_{dc,f}\Big),
	\label{eq:dmfd_current_based}
\end{equation}
\noindent where \(m_{dc}\) \((\mathrm{pu_P/pu_V})\) and \(m_i\)
\((\mathrm{pu_P/pu_A})\) are the voltage and current droop coefficients,
respectively, \(m_f\) \((\mathrm{pu_\omega/pu_P})\) is the frequency droop
coefficient, and \(\Delta v_{dc}\) and \(\Delta i_{dc,f}\) are the per-unit
DC-voltage deviation and filtered current deviation; the terminal index
\(k\) is dropped where no ambiguity arises.

The current-based form is adopted because the DC-port currents enter the
node--branch network model directly as nodal inputs, so the steady-state
relations of Section~III remain linear in the network variables without
reference to an operating point. The DMFD was originally formulated with
the filtered AC active-power deviation in place of the current
term~\cite{Mahmood_SST_2026},
 
\begin{equation}
	\Delta \omega \;=\; m_{f}\Big( m_{dc}\,\Delta v_{dc} \;-\; m_{p}\,\Delta P_{f}\Big),
	\label{eq:dmfd_power_based}
\end{equation}
where \(m_{p}\) \((\mathrm{pu_P/pu_P})\) is the power droop coefficient and
\(\Delta P_{f}\) is the per-unit AC active-power deviation about its
schedule, passed through the same first-order filter as the current term;
the sign difference with respect to
\eqref{eq:dmfd_current_based} reflects the convention
\(P_{ac}=-V_{dc}\,i_{dc}\), so both forms shift frequency in the same
direction for a given DC-link energy imbalance. The two forms differ only in
where the product \(P_{dc}=V_{dc}i_{dc}\) is linearized. With
\eqref{eq:dmfd_power_based}, the linearization enters the droop law itself
and introduces the operating-point pair \((V_{dc,0},\,i_{dc,0})\) into the
steady-state network equations. With \eqref{eq:dmfd_current_based}, the
steady-state equations are free of such terms, and in the small-signal model
of Section~IV the linearization is confined to the converter power identity
that couples the DC and AC ports. The DC-network matrices are identical for
both forms.

\paragraph*{\textbf{Power--Limit Controller}}
The nominal DMFD law adjusts the converter active-power exchange
indirectly through its frequency command. Consequently, a sustained
dc-side voltage or current deviation may continue shifting the
frequency reference after the converter reaches its current or power
capability. The fast circular current limiter bounds the commanded
current magnitude, but it does not by itself impose an active-power
limit because the terminal voltage also enters the power relation.
Continued droop action under this condition can maintain the current
references at their limits and degrade the outer voltage regulation.
A supplementary local power-limit loop is therefore included to
constrain the sustained bidirectional active-power exchange.

As shown in Fig.~\ref{fig:ctrl}, the measured power magnitude
$|P_{\mathrm{ac}}|$ is compared with the symmetric limit $P_{\max}$.
The resulting power-margin error drives a PI controller whose output is
limited at zero and can therefore take only nonpositive values. When
$|P_{\mathrm{ac}}|>P_{\max}$, the PI output becomes negative. A sign
block then assigns the required direction: the correction is negative
for positive power export and positive for negative power transfer,
thereby reducing $|P_{\mathrm{ac}}|$ in either direction. When the power
returns below the limit, the opposite error unwinds the PI state toward
zero. Clamping anti-windup holds the integrator once the output reaches
its zero limit.

The limiter output $\Delta f_{\mathrm{lim}}^{\mathrm{pu}}$ is added
after the nominal DMFD gain $m_f$ and before conversion of the per-unit
frequency reference to angular frequency. The resulting command is
\begin{equation}
	\Delta f^{\mathrm{pu}}
	=
	m_f\left(
	m_{dc}\Delta v_{dc}
	+m_i\Delta i_{dc,f}
	\right)
	+\Delta f_{\mathrm{lim}}^{\mathrm{pu}}.
	\label{eq:dmfd-power-limited}
\end{equation}
Thus, the limiting action is zero during normal operation and does not
modify the nominal DMFD dynamics. Frequency adjustment for sustained
active-power limiting is consistent with grid-forming power-limiting
droop control \cite{Iraniparast_2025}. Accordingly, the steady-state
and small-signal models in Sections~III and IV are developed for the
unsaturated condition $\Delta f_{\mathrm{lim}}^{\mathrm{pu}}=0$.

The power-limit loop complements, rather than replaces, the fast
voltage-dependent circular current limiter in Fig.~\ref{fig:ctrl}. The
current limiter constrains the commanded $dq$-current magnitude during
fast transients and faults and applies the adopted active-power
priority by preserving the active-current reference and reducing the
reactive-current reference as required. Clamping anti-windup is applied
to the outer voltage-control PI regulators while their current
references are limited.

\section{Steady-State Model}
\label{sec:ss}

This section derives the steady operating point of a meshed DC network whose
terminals apply the current-based DMFD, proves that the operating point is
unique, and specializes the result to a steady frequency bias at one terminal
and to the islanding of one terminal. The resulting maps are then used to
construct design envelopes for a four-terminal ring.

\subsection{Network Model}
\label{subsec:ss-network}

Let the node set be \(\mathcal{N}=\{1,\dots,N\}\) and the branch set
\(\mathcal{L}=\{1,\dots,N_L\}\), with a fixed orientation
\(\ell=a\!\to\!b\) on each branch. The node--branch incidence matrix is
\(M\in\mathbb{R}^{N\times N_L}\) (row \(=\) node, column \(=\) branch), with
\(+1\) at the source node \(a\), \(-1\) at the sink node \(b\), and zeros
elsewhere. Branch series parameters are \(R=\mathrm{diag}(R_\ell)\)
(\(\Omega\)) and \(L=\mathrm{diag}(L_\ell)\) (H), and node capacitors are
\(C=\mathrm{diag}(C_k)\) (F). Node voltages (pole-to-pole) are stacked in
\(V_{dc}\in\mathbb{R}^N\) (V), with steady deviations
\(\Delta V_{dc}\triangleq V_{dc}-V_{dc,\mathrm{nom}}\) (V); the
corresponding per-unit quantities are written in lower case,
\(\Delta v_{dc}\). Branch currents are stacked in
\(i_L\in\mathbb{R}^{N_L}\) (A). Converter DC-port currents
\(i_{dc}\in\mathbb{R}^N\) (A) and exogenous nodal injections
\(w\in\mathbb{R}^N\) (A) are defined positive \emph{into} the DC grid.

At steady state, KCL and Ohm's law give
\begin{equation}
	M\,i_L \;=\; i_{dc} + w,
	\label{eq:ss_kcl_ohm_ref}
\end{equation}
\begin{equation}
	i_L \;=\; R^{-1} M^\top \Delta V_{dc}.
	\label{eq:ss_ohm_idl}
\end{equation}
Eliminating \(i_L\) yields
\begin{equation}
	L_R\,\Delta V_{dc} \;=\; i_{dc} + w,
	\qquad
	L_R \;\triangleq\; M R^{-1} M^\top,
	\label{eq:ss_core_ref}
\end{equation}
where \(L_R\) is the resistive Laplacian of the DC mesh.

\subsection{DMFD Steady-State Relation and Effective Droop}
\label{sec:pu_policy}

Each converter \(k\) uses its own rating pair
\(\big(V_k^{\mathrm{u}},\,P_k^{\mathrm{u}}\big)\), with
\(I_k^{\mathrm{u}}:=P_k^{\mathrm{u}}/V_k^{\mathrm{u}}\). The diagonal
per-unit meters are
\begin{equation}
	\Gamma_V := \mathrm{diag}\!\Big(\tfrac{1}{V_k^{\mathrm{u}}}\Big),
	\qquad
	\Gamma_I := \mathrm{diag}\!\Big(\tfrac{1}{I_k^{\mathrm{u}}}\Big).
	\label{eq:pu_meters_ref}
\end{equation}
The network matrices \(M\), \(R\), and \(L_R\) remain in SI units; only the
controller inputs are scaled to per unit. The DMFD gains are tuned locally
and collected in
\begin{subequations}
	\label{eq:dmfd-perunit-gains}
	\begin{align}
		M_{dc} &= \mathrm{diag}(m_{dc,k}) \quad (\mathrm{pu_P/pu_V}),\\
		M_{i}  &= \mathrm{diag}(m_{i,k})  \quad (\mathrm{pu_P/pu_A}),\\
		M_{f}  &= \mathrm{diag}(m_{f,k})  \quad (\mathrm{pu_\omega/pu_P}),
	\end{align}
\end{subequations}
with \(m_{dc,k},\,m_{i,k},\,m_{f,k}>0\).

Let \(i_{dc}^{\mathrm{ref}}\in\mathbb{R}^N\) (A) be the scheduled DC
current, positive into the grid, and define the stacked current deviation
\begin{equation}
	\Delta i_{dc} \;\triangleq\; i_{dc} - i_{dc}^{\mathrm{ref}}.
	\label{eq:ss_delta_idc_def}
\end{equation}
At steady state the filtered current equals the measured current, so with
inactive converter limits and zero power-limit bias the DMFD relation
\eqref{eq:dmfd_current_based} reads, in stacked form,
\begin{equation}
	M_f^{-1}\Delta\omega
	\;=\; M_{dc}\Gamma_V\Delta V_{dc}
	\;+\; M_i\Gamma_I\,\Delta i_{dc},
	\label{eq:ss-dmfd}
\end{equation}
where \(\Delta\omega\in\mathbb{R}^N\) is the per-unit frequency-deviation
vector. Define the DMFD ratio of terminal \(k\),
\begin{equation}
	\rho_k \;\triangleq\; \frac{m_{dc,k}}{m_{i,k}}
	\quad (\mathrm{pu_A/pu_V}),
	\label{eq:design-rho-def}
\end{equation}
stacked as \(\rho\in\mathbb{R}^N\), the effective SI droop
\begin{equation}
	G_{\mathrm{droop}}^{\mathrm{eff}}
	\;:=\; \Gamma_I^{-1} M_i^{-1} M_{dc} \Gamma_V
	\;=\; \Gamma_I^{-1}\,\mathrm{diag}(\rho)\,\Gamma_V
	\quad (\mathrm{A/V}),
	\label{eq:pu_Geff_def_ref}
\end{equation}
and the frequency-to-current map
\begin{equation}
	B \;:=\; \Gamma_I^{-1} M_i^{-1} M_f^{-1}
	\quad (\mathrm{A/pu_\omega}).
	\label{eq:ss-B}
\end{equation}
Solving \eqref{eq:ss-dmfd} for the physical converter current gives
\begin{equation}
	i_{dc} \;=\; i_{dc}^{\mathrm{ref}}
	\;-\; G_{\mathrm{droop}}^{\mathrm{eff}}\,\Delta V_{dc}
	\;+\; B\,\Delta\omega.
	\label{eq:ss-idc-closure}
\end{equation}
Equation \eqref{eq:ss-idc-closure} is the only form in which the controller
enters the network model: the per-unit gains appear exclusively through the
SI matrices \(G_{\mathrm{droop}}^{\mathrm{eff}}\) and \(B\), which is what
allows terminals of different ratings to share one physical network model.

\subsection{Operating Point, Existence, and Uniqueness}
\label{subsec:ss-closures-dmfd}

Substituting \eqref{eq:ss-idc-closure} into \eqref{eq:ss_core_ref} yields
\begin{equation}
	\left(L_R+G_{\mathrm{droop}}^{\mathrm{eff}}\right)\Delta V_{dc}
	\;=\; i_{dc}^{\mathrm{ref}} + w + B\,\Delta\omega.
	\label{eq:ss-main}
\end{equation}

Because the DMFD acts only on the terminal frequency, the diagonal term
\(G_{\mathrm{droop}}^{\mathrm{eff}}\) in \eqref{eq:ss-main} is not imposed by
a DC-side regulator; it is a consequence of AC synchronization, which holds
\(\Delta\omega_k\) at a fixed value at every synchronized terminal. With it in
place, invertibility of the coefficient matrix follows from the standard
argument. Consider unsaturated operation, a connected DC network with
positive line resistances, and the effective droop coefficients
\(g_k:=\rho_k I_k^{\mathrm{u}}/V_k^{\mathrm{u}}\), which are strictly positive
for the positive gains used here. Setting the right-hand side of
\eqref{eq:ss-main} to zero, the equation at node \(k\) becomes
\[
\sum_{j\text{ connected to }k}
\frac{\Delta V_{dc,k}-\Delta V_{dc,j}}{R_{kj}}
+ g_k\Delta V_{dc,k} = 0 .
\]
Multiplying each nodal equation by \(\Delta V_{dc,k}\) and summing over all
nodes gives
\[
\sum_{\text{lines }(a,b)}
\frac{\left(\Delta V_{dc,a}-\Delta V_{dc,b}\right)^2}{R_{ab}}
+ \sum_{k=1}^{N} g_k\left(\Delta V_{dc,k}\right)^2 = 0 ,
\]
where each line is counted once. Every term is nonnegative, so the line
terms require equal voltage deviations at connected nodes; connectivity makes
all deviations equal, and the positive \(g_k\) force this common value to
zero. Hence zero is the only solution of the homogeneous equation,
\(L_R+G_{\mathrm{droop}}^{\mathrm{eff}}\) is invertible, and
\begin{equation}
	H \;:=\; \left(L_R+G_{\mathrm{droop}}^{\mathrm{eff}}\right)^{-1}
	\label{eq:ss-H}
\end{equation}
is well defined. The DC-voltage deviations and branch currents are then
\begin{equation}
	\Delta V_{dc} \;=\; H\left(i_{dc}^{\mathrm{ref}}+w+B\,\Delta\omega\right),
	\label{eq:ss-bias-solve_1}
\end{equation}
\begin{equation}
	i_L \;=\; R^{-1}M^\top\Delta V_{dc},
	\label{eq:ss-iL}
\end{equation}
and the converter currents follow from \eqref{eq:ss-idc-closure}. Thus each
specified frequency vector determines a unique DC-network solution for fixed
\(i_{dc}^{\mathrm{ref}}\) and \(w\).

Summing the nodal equations \eqref{eq:ss_kcl_ohm_ref} cancels all branch
currents, since each leaves one node and enters another, so
\(\sum_{k}\left(i_{dc,k}+w_k\right)=0\). Substituting
\eqref{eq:ss-idc-closure} at \(\Delta\omega=0\) gives
\(\sum_k g_k\Delta V_{dc,k}=\sum_k\big(i_{dc,k}^{\mathrm{ref}}+w_k\big)\):
if the scheduled converter currents do not balance the exogenous
injections, the droop corrections \(-g_k\Delta V_{dc,k}\) absorb the
mismatch, and the resulting DC-voltage deviations are the price of that
absorption.

At nominal AC frequency, \(\Delta\omega=0\), the nominal-frequency operating
point is
\begin{equation}
	\begin{aligned}
		\Delta V_{dc,0} &= H\left(i_{dc}^{\mathrm{ref}}+w\right),\\
		V_{dc,0} &= V_{dc,\mathrm{nom}}+\Delta V_{dc,0},
	\end{aligned}
	\label{eq:ss-pinned}
\end{equation}
\begin{equation}
	i_{dc,0} \;=\; i_{dc}^{\mathrm{ref}}
	- G_{\mathrm{droop}}^{\mathrm{eff}}\,\Delta V_{dc,0}.
	\label{eq:ss-iact}
\end{equation}
These are actual operating-point values, which in general differ from the
nominal voltage and the scheduled current.

\subsection{Frequency Bias and Islanding}
\label{subsec:ss-cases}

\subsubsection{Single-Terminal Grid-Frequency Bias}
\label{subsec:ss-grid-frequency-bias}

Suppose terminal \(k\) is synchronized to an AC grid with a prescribed steady
frequency deviation \(\Delta\omega_{g,k}\), while all other terminals remain
at nominal frequency, so that \(\Delta\omega=e_k\Delta\omega_{g,k}\) with
\(e_k\) the \(k\)-th canonical vector. The equivalent nodal-current offset is
\begin{equation}
	b \;:=\; B\,e_k\,\Delta\omega_{g,k},
	\label{eq:ss-bias-b}
\end{equation}
and \eqref{eq:ss-bias-solve_1} becomes
\begin{equation}
	\Delta V_{dc} \;=\; H\left(i_{dc}^{\mathrm{ref}}+w+b\right).
	\label{eq:ss-bias-solve}
\end{equation}
Relative to \eqref{eq:ss-pinned}, the DC-voltage shift is \(Hb\); converter
and branch currents follow from \eqref{eq:ss-idc-closure} and
\eqref{eq:ss-iL}, respectively.

\subsubsection{Islanded Terminal}
\label{subsec:ss-islanded-terminal}

When terminal \(k\) supplies an AC island, its frequency is no longer
prescribed and is determined by the island power balance. With the other
terminals at nominal frequency, \(\Delta\omega=e_k\Delta\omega_k\). Let
\(P_{\mathrm{ac},k}^{\mathrm{req}}\) be the specified,
frequency-independent net island demand, positive for power supplied to the
island. Neglecting converter losses,
\begin{equation}
	P_{\mathrm{ac},k}^{\mathrm{req}}
	\;=\; P_{\mathrm{ac},k} \;=\; -V_{dc,k}\,i_{dc,k}.
	\label{eq:ss-island-balance}
\end{equation}

For fixed \(i_{dc}^{\mathrm{ref}}\) and \(w\), the nominal-frequency
operating point \eqref{eq:ss-pinned}--\eqref{eq:ss-iact} is used as the
expansion point. Define \(\widetilde V_{dc}:=V_{dc}-V_{dc,0}\) and
\(\widetilde i_{dc}:=i_{dc}-i_{dc,0}\), so that
\(\widetilde V_{dc}=\Delta V_{dc}-\Delta V_{dc,0}\). Subtracting
\eqref{eq:ss-pinned} from \eqref{eq:ss-bias-solve_1} gives
\begin{equation}
	\widetilde V_{dc} \;=\; HB\,\Delta\omega,
	\label{eq:ss-V-affine}
\end{equation}
and subtracting \eqref{eq:ss-iact} from \eqref{eq:ss-idc-closure} and
substituting \eqref{eq:ss-V-affine} gives
\begin{equation}
	\widetilde i_{dc}
	\;=\; -G_{\mathrm{droop}}^{\mathrm{eff}}\widetilde V_{dc} + B\,\Delta\omega
	\;=\; \left(B-G_{\mathrm{droop}}^{\mathrm{eff}}HB\right)\Delta\omega.
	\label{eq:ss-idc-affine}
\end{equation}
The operating-point AC power is
\(P_{\mathrm{ac},0}=-\operatorname{diag}(V_{dc,0})\,i_{dc,0}\). Expanding
\(P_{\mathrm{ac}}=-\operatorname{diag}(V_{dc})\,i_{dc}\) and neglecting the
second-order product \(\operatorname{diag}(\widetilde V_{dc})\widetilde i_{dc}\)
yields
\begin{equation}
	P_{\mathrm{ac}}-P_{\mathrm{ac},0}
	\;\approx\; -\operatorname{diag}(V_{dc,0})\,\widetilde i_{dc}
	\;-\; \operatorname{diag}(i_{dc,0})\,\widetilde V_{dc}.
	\label{eq:ss-Pac-lin}
\end{equation}
Substituting \eqref{eq:ss-V-affine} and \eqref{eq:ss-idc-affine} into
\eqref{eq:ss-Pac-lin} gives the frequency--power map
\begin{equation}
	P_{\mathrm{ac}} \;\approx\; P_{\mathrm{ac},0} + K\,\Delta\omega,
	\label{eq:ss-Pac-affine}
\end{equation}
\begin{equation}
	\begin{aligned}
		K := {}& \operatorname{diag}(V_{dc,0})
		\left(G_{\mathrm{droop}}^{\mathrm{eff}}HB-B\right)\\
		&-\operatorname{diag}(i_{dc,0})\,HB
		\qquad (\mathrm{W}/\mathrm{pu}_{\omega}).
	\end{aligned}
	\label{eq:ss-K}
\end{equation}

Since only \(\Delta\omega_k\) is nonzero, the \(k\)-th row of
\eqref{eq:ss-Pac-affine}, together with \eqref{eq:ss-island-balance}, gives
\begin{equation}
	P_{\mathrm{ac},k}^{\mathrm{req}}
	\;\approx\; P_{\mathrm{ac},k,0} + K_{kk}\,\Delta\omega_k,
	\label{eq:ss-single-Kkk}
\end{equation}
and hence
\begin{equation}
	\Delta\omega_k \;\approx\;
	\frac{P_{\mathrm{ac},k}^{\mathrm{req}}-P_{\mathrm{ac},k,0}}{K_{kk}}
	\;=\; \frac{\Delta P_{\mathrm{ac},k}}{K_{kk}},
	\label{eq:ss-single-omega}
\end{equation}
where \(\Delta P_{\mathrm{ac},k}\) is the required change in island power.
Existence and uniqueness of the islanded operating point therefore require
\(K_{kk}\neq0\), i.e., a nonzero steady frequency-to-power sensitivity at the
islanded terminal; for the positive gains used in this paper the condition
is satisfied by \eqref{eq:ss-K}, and it is the only condition beyond
network connectivity that the islanded case adds to the synchronized case.
Substituting \(\Delta\omega=e_k\Delta\omega_k\) into
\eqref{eq:ss-bias-solve_1}, \eqref{eq:ss-idc-closure}, and
\eqref{eq:ss-iL} gives the corresponding DC voltages and currents.

\subsection{Design Envelopes for a Four-Terminal Ring}
\label{subsec:ss-design-envelopes}

\begin{figure}[t]
	\centering
	\includegraphics[width=2.4 in]{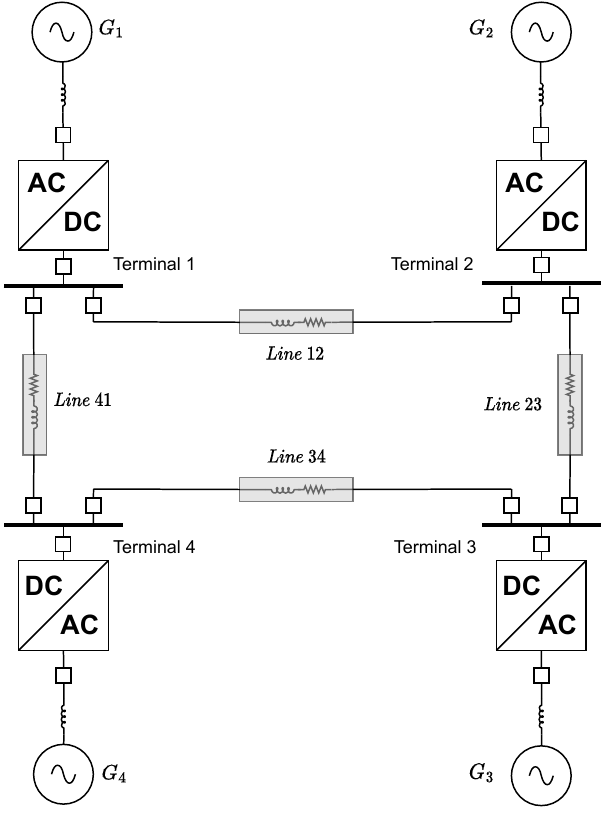}
	\caption{Simplified single-line diagram of the considered 4-terminal (ring) topology.}
	\label{fig:sys}
\end{figure}

The four-terminal ring of Fig.~\ref{fig:sys} is adopted for the remainder of
the paper; network and baseline control parameters are listed in
Table~\ref{tab:sys-params}. The node set is \(\mathcal{N}=\{1,2,3,4\}\) and
the branch set is \(\mathcal{L}=\{12,23,34,41\}\), oriented \(1\to2\),
\(2\to3\), \(3\to4\), and \(4\to1\), which fixes the incidence matrix
\(M\in\mathbb{R}^{4\times4}\). Branch currents
\(i_L=[\,i_{12},\,i_{23},\,i_{34},\,i_{41}\,]^\top\) (A) are positive along
these orientations, and the branch and node matrices are
\(R=\mathrm{diag}(R_{12},R_{23},R_{34},R_{41})\) (\(\Omega\)),
\(L=\mathrm{diag}(L_{12},L_{23},L_{34},L_{41})\) (H), and
\(C=\mathrm{diag}(C_1,C_2,C_3,C_4)\) (F); \(L\) and \(C\) are used in
Section~\ref{sec:dyn_analysis}.

Disturbances and schedules are represented by a per-unit pattern vector
\(D\in\mathbb{R}^N\), whose signs set injection or withdrawal at each node,
scaled by an amplitude \(\alpha\in[0,1]\) and mapped to amperes as
\begin{equation}
	i_{dc}^{\mathrm{ref}} + w \;=\; \Gamma_I^{-1}\,(\alpha\,D)
	\quad (\mathrm{A}).
	\label{eq:design-alpha-map}
\end{equation}
From \eqref{eq:ss-bias-solve} and \eqref{eq:ss-H}, the node-voltage
deviation satisfies
\begin{equation}
	\bigl(L_R + G_{\mathrm{droop}}^{\mathrm{eff}}\bigr)\,\Delta V_{dc}
	\;=\; \Gamma_I^{-1}\,(\alpha\,D) + b ,
	\label{eq:design-ss-solve}
\end{equation}
with \(b=0\) unless a frequency bias is present, and is reported in per unit
as \(\Delta v_{dc}=\Gamma_V\,\Delta V_{dc}\). The admissible DC-voltage band
is \(\pm\Upsilon\) with \(\Upsilon=\pm 0.05\)~pu, and
\(I_{\max}\) denotes the largest per-unit disturbance amplitude
\(\|\alpha D\|_\infty\) considered. Unless stated otherwise, the droop is
uniform, \(\rho_k=\rho\).

\subsubsection{Nominal Frequency (\(\Delta\omega=0\))}

\begin{figure}[t]
	\centering
	\includegraphics[width=\columnwidth]{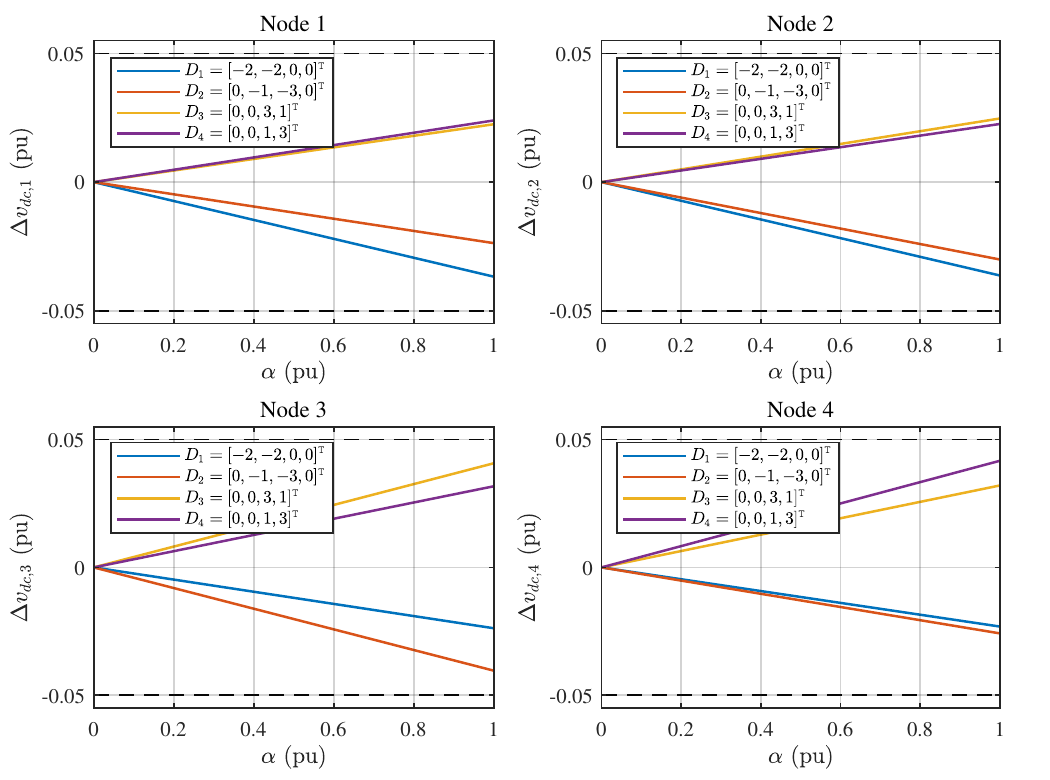}
	\caption{Node-wise \(\Delta v_{dc,k}\) (pu) versus amplitude \(\alpha\) (pu) at nominal frequency (\(\Delta\omega=0\)) and uniform droop \(\rho\) for four patterns \(D\). The dashed lines indicate the \(\pm\Upsilon\) (\(\pm 0.05 \mathrm{pu}\)) band.}
	\label{fig:A1-panels}
\end{figure}

\begin{figure}[t]
	\centering
	\subfloat[]{%
		\includegraphics[width=0.50\columnwidth]{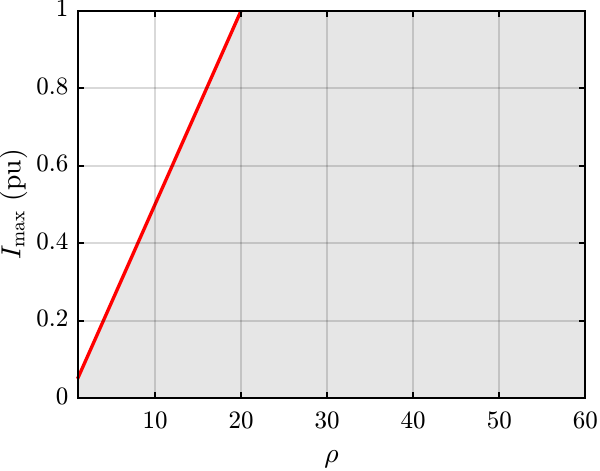}%
		\label{fig:C1-feasible}}
	\hfill
	\subfloat[]{%
		\includegraphics[width=0.50\columnwidth]{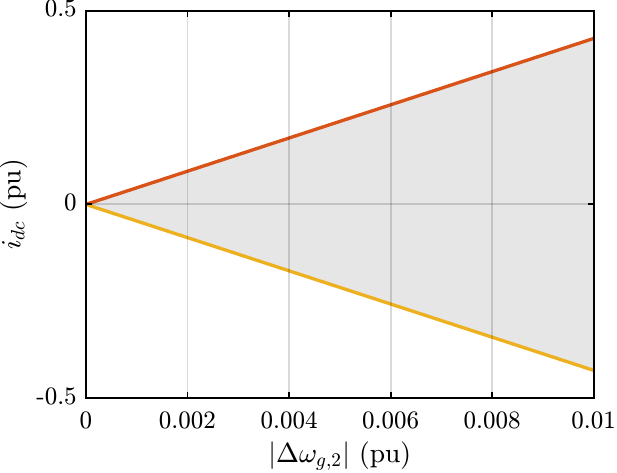}%
		\label{fig:Bp1-idc-omega}}
	\caption{Steady-state envelopes with all terminals grid-connected.
		(a) Operating-band compliance for uniform \(\rho\): contour of \(\|H_{\mathrm{pu}}(\rho)\|_\infty I_{\max}=\Upsilon\) and shaded feasible region where \(\max_k|\Delta v_{dc,k}|\le\Upsilon\).
		(b) Single-terminal frequency bias: worst-case converter current \(i_{dc}\) (pu) versus \(|\Delta\omega|\) (pu).}
	\label{fig:SS}
\end{figure}

Fig.~\ref{fig:A1-panels} reports node-wise voltage deviations under four
canonical patterns \(D\) while sweeping \(\alpha\). The near-linear slopes
predicted by
\(H_{\mathrm{pu}}(\rho):=\Gamma_V\big(L_R+G_{\mathrm{droop}}^{\mathrm{eff}}(\rho)\big)^{-1}\Gamma_I^{-1}\)
are visible, and the \(\pm\Upsilon\) band identifies the limiting node for
each pattern. Fig.~\ref{fig:C1-feasible} converts the same map into a safe
operating region in the \((\rho,I_{\max})\) plane using the induced
\(\ell_\infty\) bound
\(\max_k|\Delta v_{dc,k}|\le\|H_{\mathrm{pu}}(\rho)\|_\infty\,I_{\max}\).

\subsubsection{Single-Terminal Frequency Bias}

Fig.~\ref{fig:Bp1-idc-omega} reports the converter-current envelope under a
steady frequency bias at one terminal. The worst-case per-unit \(i_{dc}\)
magnitude grows linearly with \(|\Delta\omega|\), as predicted by
\eqref{eq:ss-idc-closure} with \(\Delta\omega=e_k\Delta\omega_{g,k}\).

\subsubsection{Single-Terminal Island}

\begin{figure}[t]
	\centering
	\subfloat[]{%
		\includegraphics[width=0.51\columnwidth]{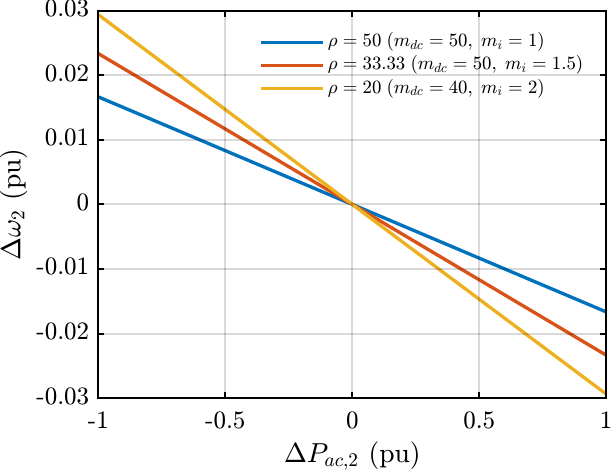}%
		\label{fig:S1a-omega}}
	\hfill
	\subfloat[]{%
		\includegraphics[width=0.49\columnwidth]{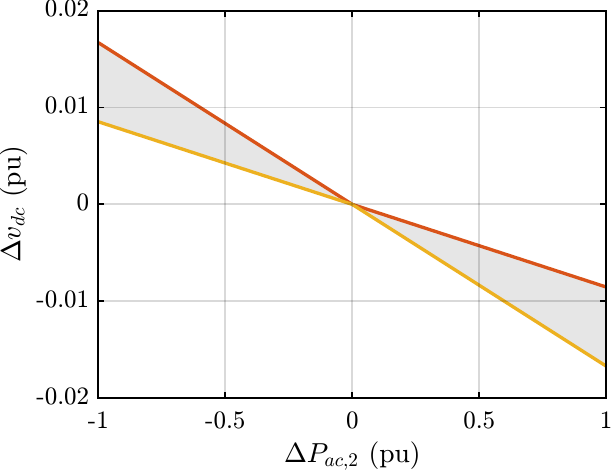}%
		\label{fig:S1b-v}}
	\caption{Steady-state envelopes for islanded Terminal~2.
		(a) Per-unit frequency deviation \(\Delta\omega_{2}\) versus per-unit island power change \(\Delta P_{\mathrm{ac},2}\).
		(b) Resulting DC-voltage excursion \(\max_k|\Delta v_{dc,k}|\) (pu) versus \(\Delta P_{\mathrm{ac},2}\) (pu).}
	\label{fig:SS_island}
\end{figure}

Fig.~\ref{fig:S1a-omega} shows the islanded frequency
\(\Delta\omega_2\) implied by \eqref{eq:ss-single-omega} versus the island
power change \(\Delta P_{\mathrm{ac},2}\) for a set of \(\rho\) values,
illustrating the frequency--power sensitivity at the islanded terminal.
Fig.~\ref{fig:S1b-v} reports the resulting \(\max_k|\Delta v_{dc,k}|\) for
the same sweep, with the \(\pm\Upsilon\) band overlaid.

\section{DC Network Dynamic Model With Current-Based DMFD Control}
\label{sec:dyn_analysis}

The model couples the DC network, DMFD, and AC active-power dynamics
with references fixed and limiting loops inactive. The AC side is
represented by a reduced-order, lossless power-angle model with fixed
voltage magnitudes, and the analysis is confined to the synchronized
grid-connected operating point; the model is intended for gain
selection and for identifying network-coupled modes, not for predicting
large-signal fault or islanding transients, which are examined by
simulation in Section~V. Network variables remain in SI units,
with per-unit controller inputs as in Section~\ref{sec:pu_policy}.

\subsection{DC-Grid Dynamic Equations}

Using the network orientation defined in Section~III, the DC-node
voltages and branch currents satisfy
\begin{align}
    C\dot V_{dc} &=-Mi_L+i_{dc}+w, \label{eq:KCL_exact}\\
    L\dot i_L &=M^\top V_{dc}-Ri_L. \label{eq:KVL_exact}
\end{align}

\subsection{Operating Point and Small-Signal Deviations}

Let \((V_{dc,0},i_{L,0},i_{dc,0},w_0)\) be operating-point values
satisfying the steady network and controller equations, i.e., \eqref{eq:ss-pinned}--\eqref{eq:ss-iact}, with
\begin{equation}
    0=-Mi_{L,0}+i_{dc,0}+w_0,\quad
    0=M^\top V_{dc,0}-Ri_{L,0}.
    \label{eq:op_point}
\end{equation}
As in Section~III, tildes denote operating-point perturbations:
\begin{equation}
    \begin{aligned}
        \widetilde V_{dc}&:=V_{dc}-V_{dc,0},&
        \widetilde i_L&:=i_L-i_{L,0},\\
        \widetilde i_{dc}&:=i_{dc}-i_{dc,0},&
        \widetilde w&:=w-w_0.
    \end{aligned}
    \label{eq:devs_dc}
\end{equation}
In particular, \(\widetilde V_{dc}=\Delta V_{dc}-\Delta V_{dc,0}\)
and \(\widetilde i_{dc}=\Delta i_{dc}-\Delta i_{dc,0}\). Subtracting \eqref{eq:op_point} from
\eqref{eq:KCL_exact}--\eqref{eq:KVL_exact} gives
\begin{align}
    C\dot{\widetilde V}_{dc}
    &=-M\widetilde i_L+\widetilde i_{dc}+\widetilde w,
    \label{eq:KCL_dev}\\
    L\dot{\widetilde i}_L
    &=M^\top\widetilde V_{dc}-R\widetilde i_L.
    \label{eq:KVL_dev}
\end{align}

\subsection{Current-Based DMFD and Current Metering}

For a fixed current schedule, the filtered-error perturbation
\(\widetilde i_{dc,f}\) (A) satisfies
\begin{equation}
    T\dot{\widetilde i}_{dc,f}
    =\widetilde i_{dc}-\widetilde i_{dc,f},
    \qquad T=\operatorname{diag}(T_k)>0.
    \label{eq:metering}
\end{equation}
Define the per-unit frequency perturbations
\(\widetilde\omega:=\Delta\omega-\Delta\omega_0\) and
\(\widetilde\omega_g:=\Delta\omega_g-\Delta\omega_{g,0}\).
Subtracting the operating-point DMFD relation gives
\begin{equation}
    \widetilde\omega
    =M_f\left(M_{dc}\Gamma_V\widetilde V_{dc}
    +M_i\Gamma_I \widetilde i_{dc,f}\right).
    \label{eq:DMFD}
\end{equation}

\subsection{AC-Side Active-Power Map and Linearization}

The AC active-power channel is represented by the lossless
power-angle model
\begin{equation}
    P_{ac,k}=\frac{V_kV_{g,k}}{X_k}\sin\theta_k,
    \qquad\theta_k:=\delta_k-\delta_{g,k},
    \label{eq:P_nl}
\end{equation}
where \(V_k\) and \(V_{g,k}\) are the converter and local-grid
line-to-line RMS voltages, \(X_k\) is the per-phase tie reactance,
and angles are in radians. With
\(\widetilde P_{ac,k}:=P_{ac,k}-P_{ac,k,0}\), the first-order
expansion about \((V_{k,0},V_{g,k,0},\theta_{k,0})\) is
\begin{align}
    \widetilde P_{ac,k}={}&
    \underbrace{\frac{V_{g,k,0}}{X_k}\sin\theta_{k,0}}_{K_{PV_k,k}}
    \widetilde V_k
    +\underbrace{\frac{V_{k,0}}{X_k}\sin\theta_{k,0}}_{K_{PV_g,k}}
    \widetilde V_{g,k}\nonumber\\
    &+\underbrace{\frac{V_{k,0}V_{g,k,0}}{X_k}\cos\theta_{k,0}}_{K_{P\delta,k}}
    \left(\widetilde\delta_k-\widetilde\delta_{g,k}\right).
    \label{eq:P_lin}
\end{align}
Holding the AC-voltage magnitudes fixed in the reduced active-power
model eliminates the first two terms. Since
\(\dot{\widetilde\delta}_k-\dot{\widetilde\delta}_{g,k}
=\omega_{\mathrm{nom}}(\widetilde\omega_k-\widetilde\omega_{g,k})\),
where \(\omega_{\mathrm{nom}}=2\pi f_{\mathrm{nom}}\),
\begin{equation}
    \dot{\widetilde P}_{ac,k}
    =\omega_{\mathrm{nom}}K_{P\delta,k}
    \left(\widetilde\omega_k-\widetilde\omega_{g,k}\right).
    \label{eq:P_rate}
\end{equation}
The vector form uses
\(K_{P\delta}:=\operatorname{diag}(K_{P\delta,k})\) (W/rad).
A local grid held at its operating frequency has
\(\widetilde\omega_{g,k}=0\).

\subsection{DC-Port Power Identity and Current Deviation}

At the DC port,
\begin{equation}
    P_{dc,k}=V_{dc,k}i_{dc,k}.
    \label{eq:Pidc_nl}
\end{equation}
With lossless converter coupling,
\(P_{dc,k}=-P_{ac,k}\). Linearization gives
\begin{equation}
    -\widetilde P_{ac,k}
    =i_{dc,k,0}\widetilde V_{dc,k}
    +V_{dc,k,0}\widetilde i_{dc,k}.
    \label{eq:Pidc_lin}
\end{equation}
Using \(i_{dc,k,0}=-P_{ac,k,0}/V_{dc,k,0}\),
\begin{equation}
    \widetilde i_{dc,k}
    =-\frac{\widetilde P_{ac,k}}{V_{dc,k,0}}
    +\frac{P_{ac,k,0}}{V_{dc,k,0}^2}\widetilde V_{dc,k}.
    \label{eq:idc_from_P}
\end{equation}
The diagonal gains \(K_\alpha:=\operatorname{diag}(1/V_{dc,k,0})\)
and \(K_\times:=\operatorname{diag}(P_{ac,k,0}/V_{dc,k,0}^2)\),
in A/W and A/V respectively, give
\begin{equation}
    \widetilde i_{dc}
    =-K_\alpha\widetilde P_{ac}+K_\times\widetilde V_{dc}.
    \label{eq:idc_stack}
\end{equation}

\subsection{State-Space Stacking}

Substituting \eqref{eq:idc_stack} into \eqref{eq:KCL_dev} and
\eqref{eq:metering}, and \eqref{eq:DMFD} into \eqref{eq:P_rate},
closes the model. Stack
\(x=[\,\widetilde V_{dc};\widetilde i_L;
\widetilde P_{ac};\widetilde i_{dc,f}\,]
\in\mathbb{R}^{3N+N_L}\) and
\(u=[\,\widetilde w;\widetilde\omega_g\,]
\in\mathbb{R}^{2N}\). Then \(\dot x=Ax+B_u u\), with
\begin{subequations}\label{eq:N_AB}
    \begingroup\setlength{\arraycolsep}{3pt}
    \begin{align}
        A&=\begin{bmatrix}
            C^{-1}K_\times &-C^{-1}M &-C^{-1}K_\alpha &0\\
            L^{-1}M^\top &-L^{-1}R &0 &0\\
            A_{31} &0 &0 &A_{34}\\
            T^{-1}K_\times &0 &-T^{-1}K_\alpha &-T^{-1}
        \end{bmatrix},\\[3pt]
        B_u&=\begin{bmatrix}
            C^{-1}&0\\
            0&0\\
            0&-\omega_{\mathrm{nom}}K_{P\delta}\\
            0&0
        \end{bmatrix}.
    \end{align}
    \endgroup
\end{subequations}
The third-row blocks follow from \eqref{eq:DMFD} and \eqref{eq:P_rate}:
\begin{align*}
    A_{31}&=\omega_{\mathrm{nom}}K_{P\delta}M_fM_{dc}\Gamma_V,\\
    A_{34}&=\omega_{\mathrm{nom}}K_{P\delta}M_fM_i\Gamma_I.
\end{align*}

\begin{figure}[t]
	\centering
	\subfloat []{
			\includegraphics[width=0.6\columnwidth]{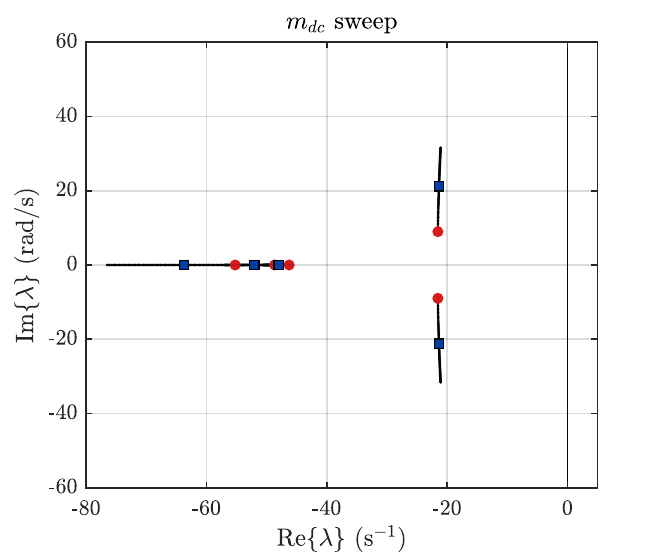}%
			\label{fig:D1}}
		\hfill
		\subfloat []{
			\includegraphics[width=0.6\columnwidth]{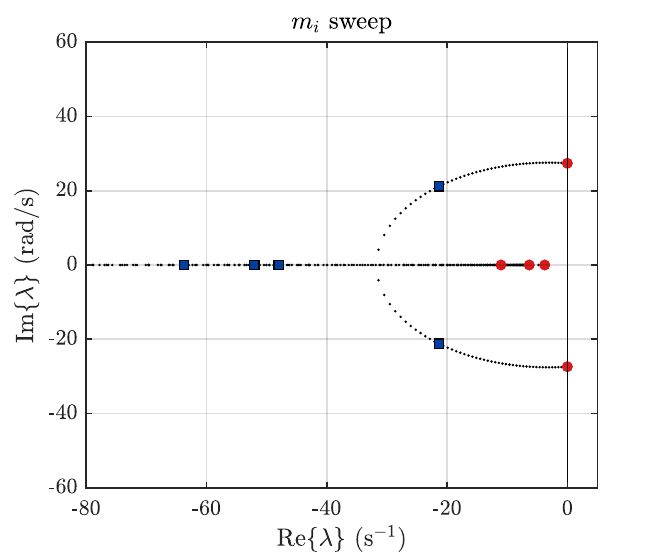}%
			\label{fig:D2}}
        \caption{Dominant eigenvalue loci under DMFD gain sweeps for the four-terminal DC ring. (a) \(m_{dc}\in[30,80]\) with \(m_i=1.5\). (b) \(m_i\in[0,5]\) with \(m_{dc}=50\). In both cases, \(m_f=0.01\). Sweep starting points are marked by filled circles and baseline poles by filled squares.}
		\label{fig:DE}
	\end{figure}
	
	\begin{figure}[t]
		\centering
		\includegraphics[width=2.7 in]{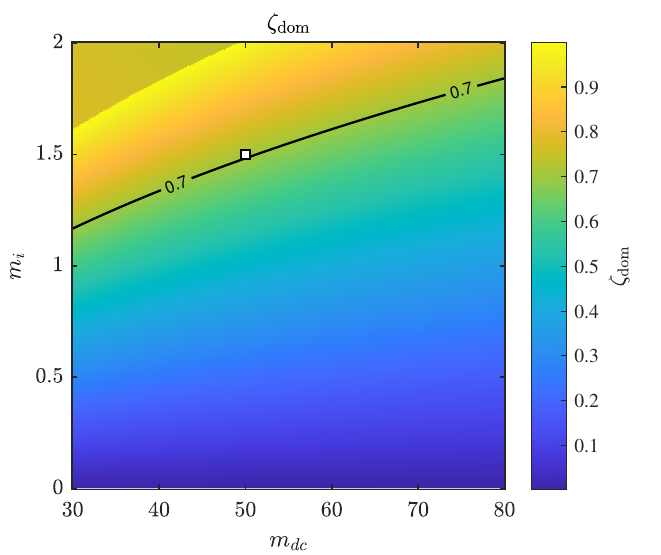}
        \caption{Damping ratio $\zeta_{\mathrm{dom}}$ of the dominant oscillatory mode over the $(m_{dc},m_i)$ gain plane at $m_f=0.01$. The baseline point $(m_{dc},m_i)=(50,1.5)$ is marked by a square, and the contour denotes $\zeta_{\mathrm{dom}}=0.7$.}
        \label{fig:D4-zeta}
	\end{figure}

\subsection{Small-Signal Dynamic Analysis Results}
\label{subsec:dyn-analysis-results}

At the baseline point, the linearized system is stable, with the
dominant oscillatory mode having a damping ratio of
$\zeta_{\mathrm{dom}}=0.71$. No $j\omega$-axis crossing occurs
over the evaluated $m_{dc}$ range in Fig.~\ref{fig:D1}. In the
$m_i$ sweep of Fig.~\ref{fig:D2}, the only marginal case occurs at
the endpoint $m_i=0$, where a conjugate pole pair lies at
$\lambda \approx \pm j27.4~\mathrm{s}^{-1}$
($4.36~\mathrm{Hz}$). For positive values of $m_i$ within the
evaluated range, this mode moves into the left-half plane. Thus, the
DC-current droop component provides damping to a mode that is undamped, leaving the system marginally stable, when the current-feedback channel is removed, highlighting
the role of the second DMFD measurement in the small-signal dynamics.

Fig.~\ref{fig:D4-zeta} further summarizes the damping ratio of the
dominant oscillatory mode over the $(m_{dc},m_i)$ gain plane,
illustrating the combined influence of the voltage- and current-droop
gains on the dominant-mode damping.

\begin{table}[t]
	\caption{System Parameters (4‑Terminal Ring: Baseline Values and Symbols)}
	\label{tab:sys-params}
	\centering
	\footnotesize
	\setlength{\tabcolsep}{5pt}
	\renewcommand{\arraystretch}{1.15}
	\begin{tabular}{@{}lll@{}}
		\toprule
		\textbf{Symbol} & \textbf{Description} & \textbf{Baseline} \\
		\midrule
		\multicolumn{3}{@{}l}{\emph{Electrical ratings and constants}} \\
		\midrule
		$f_{\mathrm{nom}}$, $\omega_{\mathrm{nom}}$ & Nominal AC frequency, rad/s & $60$ (Hz), $2\pi\!\cdot\!60$ (rad/s) \\
		$V^{\mathrm{nom}}_{dc}$ & Nominal DC (pole‑to‑pole) & $40\,\mathrm{kV}$ \\
		$V_{g,\mathrm{LL}}$ & AC grid line‑to‑line RMS & $23\,\mathrm{kV}$ \\
		$(P^{\mathrm u}_k,\; V^{\mathrm u}_k)$ & Power and voltage bases & $(20\,\mathrm{MVA},\; 40\,\mathrm{kV})$ \\
		\midrule
		\multicolumn{3}{@{}l}{\emph{DC network (4‑terminal ring)}} \\
		\midrule
		$R_\ell$ & line resistances ($\Omega$) & $[0.75,\ 0.6,\ 0.9,\ 1.5]$  \\
		$L_\ell$ & line inductances (mH) & $[4,\ 6.36,\ 4.77,\ 7.96]$  \\
		$C_k$ & Node capacitors (each) & $15$ (mF) \\
		\midrule
		\multicolumn{3}{@{}l}{\emph{AC side}} \\
		\midrule
		$R_{\mathrm{ac},k}$ & AC line/Thévenin resistance & \textit{1.35} ($\Omega$) \\
		$X_{\mathrm{th},k}$ & AC coupling reactance & \textit{4.154} ($\Omega$) \\
		\midrule
		\multicolumn{3}{@{}l}{\emph{DMFD (per‑unit gains; plant in SI)}} \\
		\midrule
		$(m_{dc},\; m_i,\; m_f)$ & DMFD gains (per unit) & $\big(50,\; 1.5,\; 0.01\big)$ \\
		$m_Q$ & Volt-var droop gain (per unit) & \textit{0.05} (pu$\mathrm{_V}$/pu$\mathrm{_Q}$) \\
		$T_k$ & Current‑meter time constant & $0.004$ (s) \\
		\midrule
		\multicolumn{3}{@{}l}{\emph{Power‑limit controller}} \\
		\midrule
		$K_{p,\ell}$ & PI proportional gain & \textit{0.025} (pu$\mathrm{_f}$/pu$\mathrm{_P}$)\\
		$K_{i,\ell}$ & PI integral gain & \textit{0.25} (pu$\mathrm{_f}$/(pu$\mathrm{_P}$$\cdot$s))\\
		 \bottomrule
	\end{tabular}

\end{table}

\section{Simulation Results and Discussion}\label{sec:results}

An averaged electromagnetic–transient (EMT) model of the proposed DMFD control is implemented in \textsc{MATLAB}/Simulink for the four‑terminal meshed DC network of Fig.~\ref{fig:sys}. Key electrical parameters and controller gains appear in Table~\ref{tab:sys-params}. 

The grid-following (GFL) baseline shares the plant, inner current
controller, and circular current limiter of the DMFD case and differs only
in synchronization and in the outer loop. It synchronizes through a
synchronous-reference-frame phase-locked loop; its active-current
reference \(i^{\mathrm{ref}}_d\) is generated by a PI controller acting on
the same DC-measurement combination
\(m_{dc}\Delta v_{dc}+m_i\Delta i_{dc,f}\), and its reactive-current
reference \(i^{\mathrm{ref}}_q\) by the same volt-var droop used in the
DMFD controller. Both controllers therefore use the common gain set
\((m_{dc},m_i,m_Q)\) in all scenarios.

\begin{figure}[t]
	\centering
	\subfloat []{
			\includegraphics[width=0.95\columnwidth]{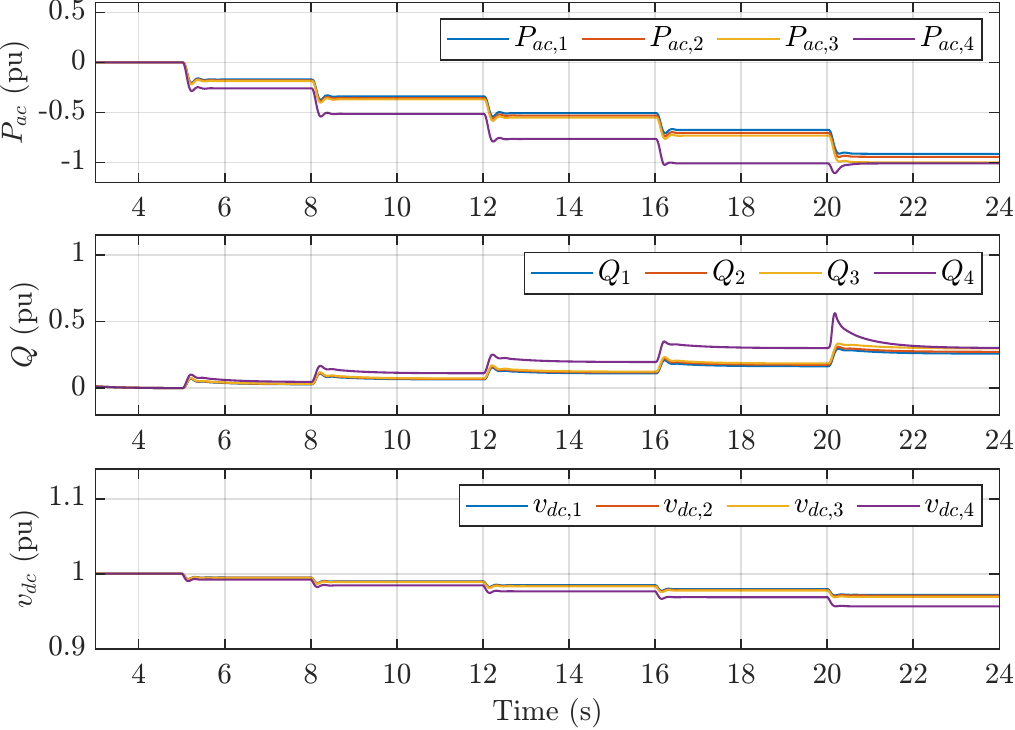}%
			\label{fig:Loading_DMFD_1.pdf}}\\
		\subfloat []{
			\includegraphics[width=0.95\columnwidth]{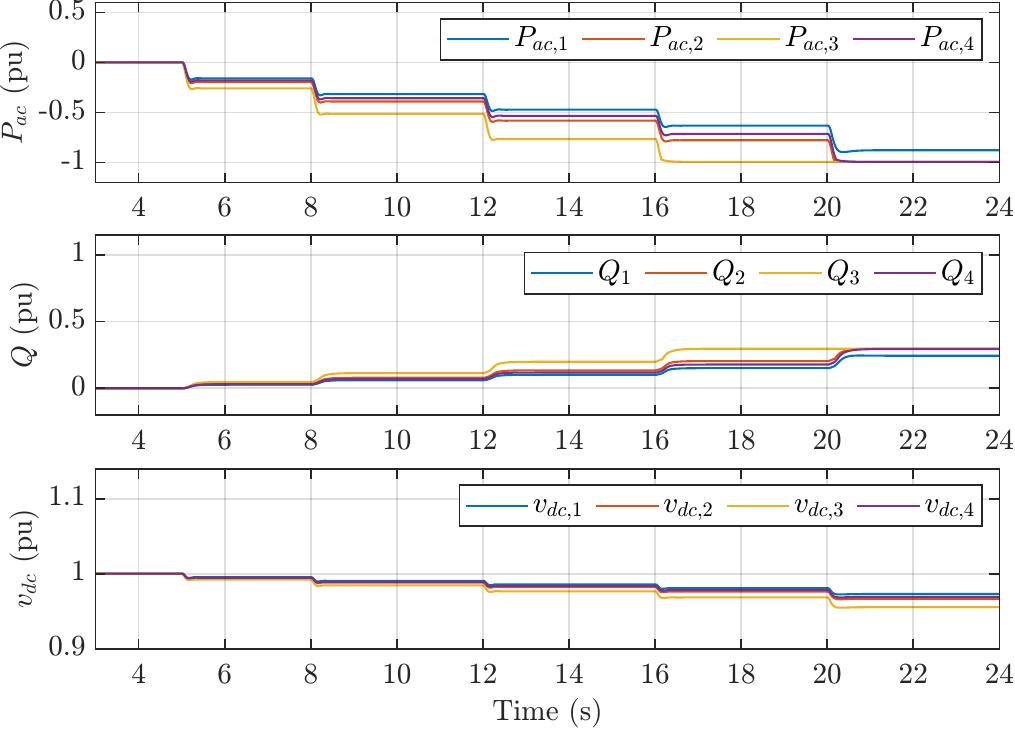}%
			\label{fig:Loading_GFL_1.pdf}}
		\caption{Loading ramps at Terminals 2 and 3: under DMFD and GFL. (a) DMFD control. (b) Conventional GFL-based control.
		}
		\label{fig:dyn_Loading}
	\end{figure}

\emph{Scenario~1: Loading and Scheduling:}
Controllable DC loads connected at terminals~2 and~3 are increased, in Fig.~\ref{fig:Loading_DMFD_1.pdf}, from \(0\) to \(1\)~pu and from \(0\) to \(3\)~pu, respectively. 
For comparison, the GFL baseline described above is simulated under the same conditions in Fig.~\ref{fig:Loading_GFL_1.pdf}.
Transient trajectories differ noticeably between DMFD and GFL, whereas the steady‑state operating points are essentially identical; for example, Terminal‑3 DC voltage settles to \(v_{dc,3}=0.9563\)~pu with DMFD and \(v_{dc,3}=0.9558\)~pu with GFL. All node voltages remain within the \(\pm\Upsilon\) band throughout both ramps, consistent with the steady-state envelopes of Fig.~\ref{fig:A1-panels}.

Small excursions outside the DC‑voltage band observed in Fig.~\ref{fig:dyn_Loading} occur when a converter reaches its rated power and power-limiting schemes are engaged. In that regime the local DC terminal voltage is reduced to enforce the limit, explicitly by the DMFD power‑limit controller and indirectly in the GFL implementation through inner‑loop saturation. Such excursions can be mitigated by design via lower DC branch resistances; scheduling alone may be insufficient in extreme conditions where several terminals operate near rating. 

\begin{figure}[t]
	\centering
	\includegraphics[width=3.4 in]{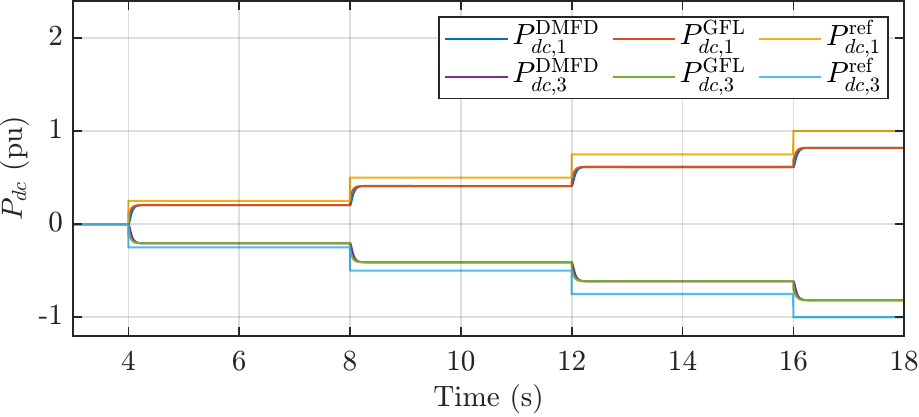}
	\caption {Power scheduling performance of DMFD control vs. conventional GFL-based control.}
	\label{fig:scheduling}
\end{figure}

Reference‑tracking was also examined by scheduling a power transfer from terminal~1 to terminal~3 (Fig.~\ref{fig:scheduling}). Both DMFD and GFL follow the commanded transfer with comparable errors. The steady‑state offset is not a DMFD artifact; it reflects the unequal line resistances and associated DC voltage drops that are likewise present under GFL for the same topology.

\begin{figure}[!t]
	\centering
	\includegraphics[width=0.95\columnwidth]{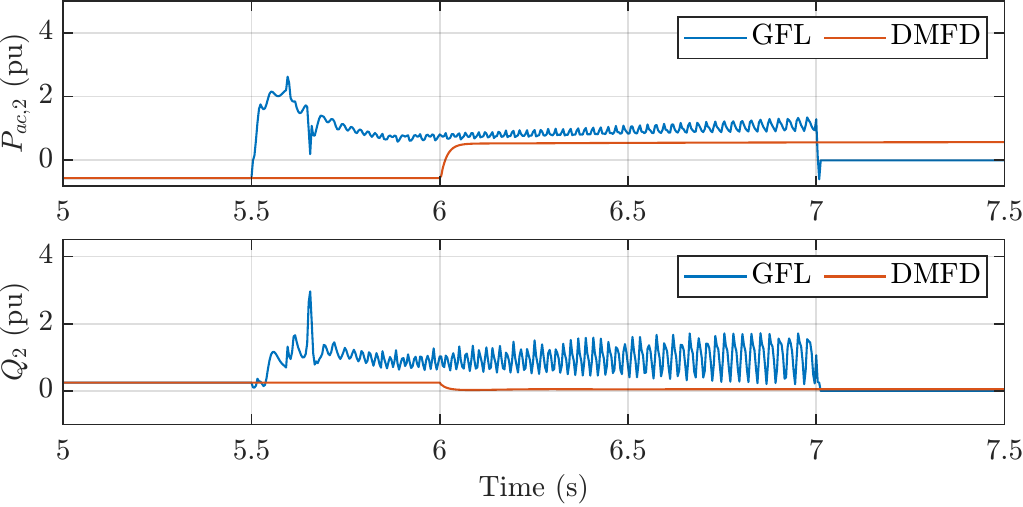}
	\caption{Islanding transition at Terminal~2 with local AC loads.
		The local grid is disconnected at $t=6$~s. Shown are
		$P_{\mathrm{ac},2}$ and $Q_2$ under DMFD and the GFL baseline.
		DMFD maintains stable operation following islanding, whereas the
		GFL baseline loses stable operation.}
	\label{fig:island_event}
\end{figure}

\emph{Scenario~2: Islanded Operation at Terminal~2:}
Islanding at Terminal~2 is evaluated through two tests. A resistive load of \(0.58\)~pu and an electronic load of \(0.10\)~pu are connected at the AC side of Terminal~2. In the first test, the local grid is disconnected at \(t=6\)~s, and the proposed DMFD controller is compared with the GFL baseline described above. As shown in Fig.~\ref{fig:island_event}, DMFD maintains stable operation through the transition from grid-connected operation to supplying the local AC island, whereas the GFL baseline loses stable operation following disconnection.

\begin{figure}[!t]
	\centering
	
	\subfloat[]%
	{\includegraphics[width=0.95\columnwidth]{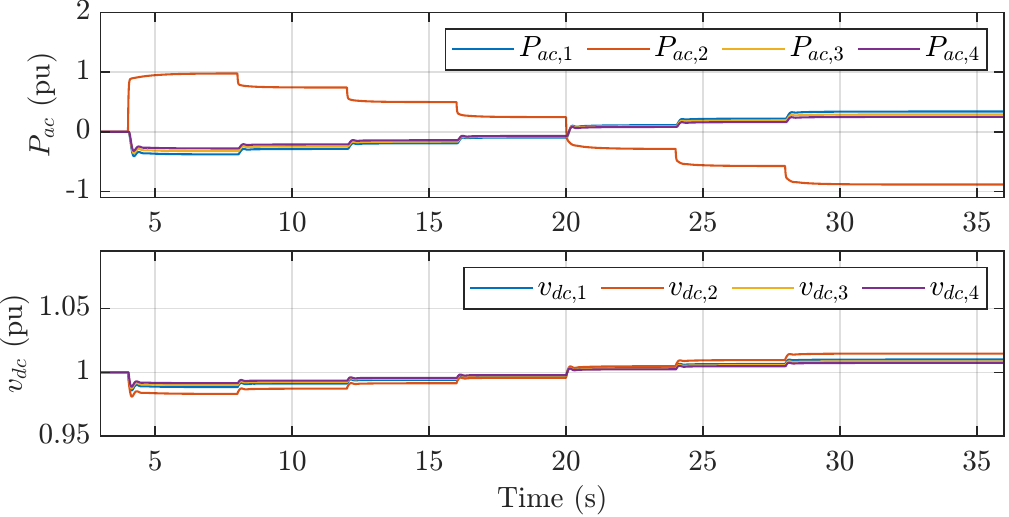}
		\label{fig:island_steps}}\\[2pt]
	
	\subfloat[]%
	{\includegraphics[width=0.48\columnwidth]{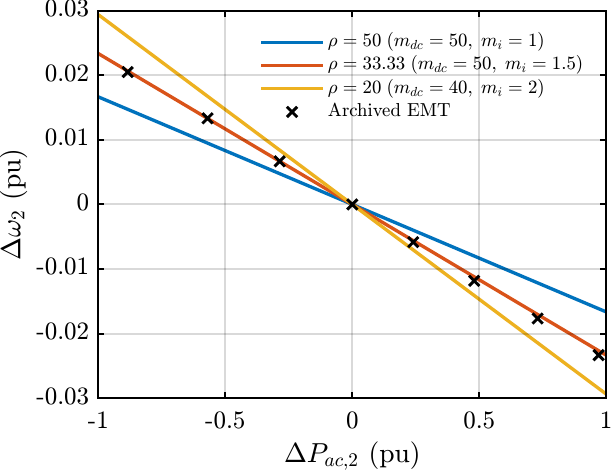}
		\label{fig:island_ss_freq}}
	\hfill
	\subfloat[]%
	{\includegraphics[width=0.48\columnwidth]{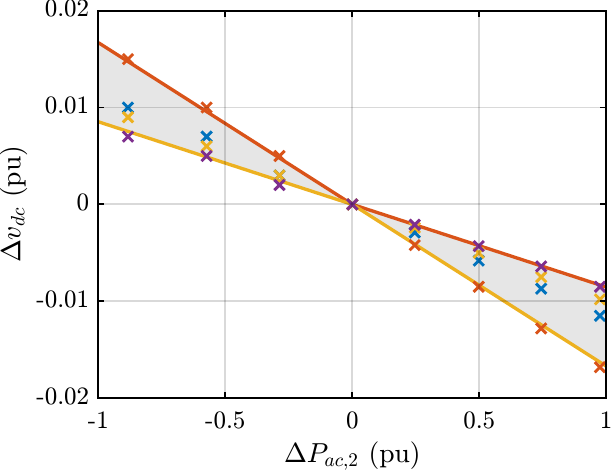}
		\label{fig:island_ss_vdc}}
	
	\caption{Islanded operation under DMFD and comparison with the
		steady-state analysis.
		(a) Stepped island power at Terminal~2, showing
		$P_{\mathrm{ac},k}$ and $v_{dc,k}$ for all terminals.
		(b) Analytical steady-state $\Delta\omega_2$ versus
		$\Delta P_{\mathrm{ac},2}$ with simulation operating points
		extracted from (a).
		(c) Analytical $\Delta v_{dc}$ envelopes versus
		$\Delta P_{\mathrm{ac},2}$ with the corresponding simulation
		operating points.}
	\label{fig:island_validation}
\end{figure}

In the second test, the island power demand is varied in steps under DMFD control to validate the steady-state islanding characteristics derived in Section~III. Figure~\ref{fig:island_validation}\subref{fig:island_steps} shows the resulting terminal active powers and DC voltages. Steady operating points extracted from this simulation are compared with the analytical results in Fig.~\ref{fig:island_validation}\subref{fig:island_ss_freq} and \subref{fig:island_ss_vdc}. The simulated \(\Delta\omega_2\)--\(\Delta P_{\mathrm{ac},2}\) points follow the predicted frequency--power characteristic, while the corresponding DC-voltage operating points remain within the analytical \(\Delta v_{dc}\) envelope. These results corroborate the steady-state model under single-terminal islanded operation.

\begin{figure}[t]
	\centering
	\includegraphics[width=3.4 in]{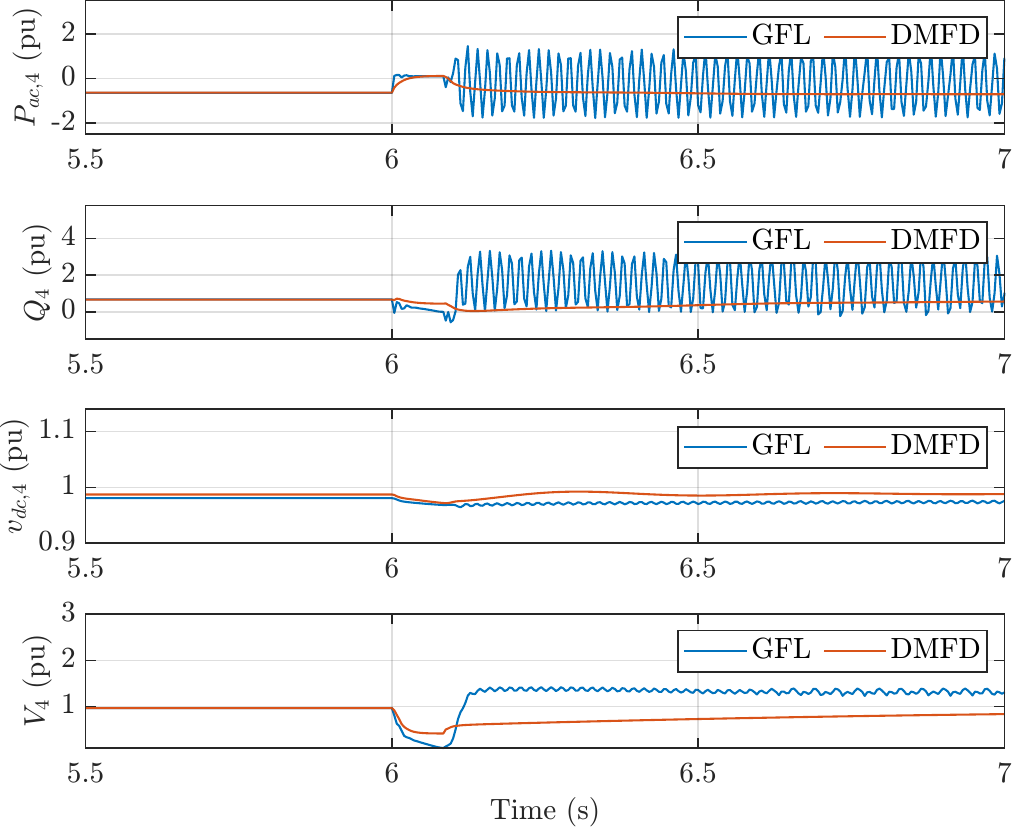}
	\caption{Fault on weak AC tie at Terminal~4 (SCR\,=\,2). 
	}
\label{fig:Fault}
\end{figure}

\emph{Scenario~3: Fault on a Weak AC Tie at Terminal~4:}
Fault ride-through was assessed with Terminal~4 connected to a weak AC tie.
For this scenario the Th\'{e}venin impedance of the Terminal-4 tie is set to
\(R_{\mathrm{ac},4}=\mathrm{4.1}~\Omega\) and
\(X_{\mathrm{th},4}=\mathrm{12.6}~\Omega\), which corresponds to a
short-circuit ratio \(\mathrm{SCR}=2\) on the \(20\)-MVA terminal base; the
remaining terminals retain the values of Table~\ref{tab:sys-params}.
A three‑phase fault is applied at the midpoint of the line at \(t=6\)~s and cleared after five 60‑Hz cycles; the terminal was pre‑loaded at \(S_4=0.75\)~pu. The test was executed for the proposed DMFD controller and for the GFL baseline described above. The results in Fig.~\ref{fig:Fault} demonstrate stable operation and post‑fault recovery with the DMFD control, whereas the GFL baseline becomes unstable under the same disturbance. These results are consistent with the improved robustness of grid‑forming behavior on weak AC ties, and illustrate the value of utilizing DMFD control in future multi-terminal DC networks.

\section{Conclusion}\label{sec:Conc}
The results of this paper establish that grid-forming behavior at the AC
port of every terminal is compatible with a well-defined operating point of
a meshed MVDC network, and that the compatibility is structural rather than
a matter of tuning: because AC synchronization alone closes an effective DC
droop under the DMFD, no terminal needs a dedicated DC-voltage regulator for
the mesh to settle, and the concern that dual-sided grid forming leaves the
DC link floating does not arise. A consequence of placing frequency in the
droop law is that an AC-frequency bias at one terminal becomes a DC-network
quantity and redistributes power across the mesh, a coupling absent in
GFL-based MTDC schemes. The analysis shows this coupling to be linear and
bounded, so it can be budgeted in the DC-voltage and current envelopes at
the design stage rather than discovered in operation. The same mechanism
allows a terminal to island without reconfiguration, with the island
frequency fixed by local power balance and the resulting DC excursion known
in advance from the island demand. On the dynamic side, the current channel
of the DMFD is not an optional refinement: its removal renders the system marginally stable, with an undamped oscillatory mode, so voltage-only droop at grid-forming terminals carries a 
stability risk that the second measurement removes. Finally, the weak-tie
results indicate that the AC-side benefit of grid forming, including fault
ride-through where a PLL-based baseline loses stability, is obtained without
giving up the DC-side behavior of conventional droop. Sensitivity of sharing
to non-uniform droop and line asymmetry, and comparison against grid-forming
terminals with voltage-only DC droop, are left for future work.

%
	\bibliographystyle{IEEEtran}
	\bibliography{references}

@techreport{nerc_2021_gfm_whitepaper,
  author      = {{NERC Reliability and Security Technical Committee (RSTC)}},
  title       = {White Paper: Grid Forming Functional Specifications for Bulk Power System Applications},
  institution = {North American Electric Reliability Corporation},
  year        = {2023},
  url         = {https://www.nerc.com/comm/RSTC_Reliability_Guidelines/White_Paper_GFM_Functional_Specification.pdf},
  note        = {Accessed 2025-09-04}
}

@article{rosso_ojia_2021_gfm_review,
  author   = {R. Rosso and X. Wang and M. Liserre and X. Lu and S. Engelken},
  title    = {Grid-Forming Converters: Control Approaches, Grid-Synchronization, and Future Trends—A Review},
  journal  = {IEEE Open Journal of Industry Applications},
  volume   = {2},
  pages    = {93--109},
  year     = {2021},
  doi      = {10.1109/OJIA.2021.3074028}
}

@article{bahrani_pesmag_2024_gfm_landscape,
  author   = {B. Bahrani and M. H. Ravanji and B. Kroposki and D. Ramasubramanian and X. Guillaud and T. Pr{\'e}vost and N.-A. Cutululis},
  title    = {Grid-Forming Inverter-Based Resource Research Landscape: Understanding the Key Assets for Renewable-Rich Power Systems},
  journal  = {IEEE Power \& Energy Magazine},
  volume   = {22},
  number   = {2},
  pages    = {18--29},
  year     = {2024},
  doi      = {10.1109/MPE.2023.3343338}
}

@article{wang_ojia_2020_sync,
  author   = {X. Wang and M. G. Taul and H. Wu and Y. Liao and F. Blaabjerg and L. Harnefors},
  title    = {Grid-Synchronization Stability of Converter-Based Resources—An Overview},
  journal  = {IEEE Open Journal of Industry Applications},
  volume   = {1},
  pages    = {115--134},
  year     = {2020},
  doi      = {10.1109/OJIA.2020.3020392}
}

@article{taul_tpel_2019_sync_assessment,
  author   = {M. G. Taul and X. Wang and P. Davari and F. Blaabjerg},
  title    = {An Overview of Assessment Methods for Synchronization Stability of Grid-Connected Converters Under Severe Symmetrical Grid Faults},
  journal  = {IEEE Transactions on Power Electronics},
  volume   = {34},
  number   = {10},
  pages    = {9655--9670},
  year     = {2019},
  doi      = {10.1109/TPEL.2019.2892142}
}

@techreport{nerc_2025_large_loads,
  author      = {{North American Electric Reliability Corporation (NERC)}},
  title       = {Characteristics and Risks of Emerging Large Loads},
  type        = {White Paper},
  institution = {North American Electric Reliability Corporation},
  month       = jul,
  year        = {2025},
  url         = {https://www.nerc.com/comm/RSTCReviewItems/3_Doc_White%20Paper%20Characteristics%20and%20Risks%20of%20Emerging%20Large%20Loads.pdf}
}

@article{Sun2011_Impedance_TPEL,
  author  = {J. Sun},
  title   = {Impedance-Based Stability Criterion for Grid-Connected Inverters},
  journal = {IEEE Transactions on Power Electronics},
  volume  = {26},
  number  = {11},
  pages   = {3075--3078},
  year    = {2011}
}

@techreport{interopera_2024_d22,
  author      = {{InterOPERA Consortium}},
  title       = {Grid-Forming Functional Requirements for {HVDC} Converter Stations and {DC}-Connected Power Park Modules in Multi-Terminal Multi-Vendor {HVDC} Systems (Deliverable D2.2)},
  institution = {EU InterOPERA Project},
  year        = {2024},
  month       = {Jan.},
  url         = {https://interopera.eu/wp-content/uploads/files/deliverables/InterOPERA-D2.2-GFM-functional-requirements-202401.pdf},
  note        = {Accessed 2025-09-04}
}

@techreport{entsoe_2019_hvdc_ops,
  author      = {{ENTSO-E}},
  title       = {HVDC Links in System Operations—Technical Paper},
  institution = {European Network of Transmission System Operators for Electricity},
  year        = {2019},
  month       = {Dec.},
  url         = {https://www.entsoe.eu/Documents/SOC%20documents/20191203_HVDC%20links%20in%20system%20operations.pdf},
  note        = {Accessed 2025-09-04}
}

@techreport{entsoe_2021_mfs_hvdc,
  author      = {{ENTSO-E}},
  title       = {Operational Limits and Conditions for Mutual Frequency Support over {HVDC}},
  institution = {European Network of Transmission System Operators for Electricity},
  year        = {2021},
  url         = {https://www.entsoe.eu/Documents/SOC%20documents/Operational_Limits_and_Conditions_for_Mutual_Frequency_Support_over_HVDC_Report.pdf},
  note        = {Accessed 2025-09-04}
}

@article{PrietoAraujo2011_MTDC_Droop_TPWRD,
  author  = {E. Prieto-Araujo and F. D. Bianchi and A. Junyent-Ferr{\'e} and O. Gomis-Bellmunt},
  title   = {Methodology for Droop Control Dynamic Analysis of Multiterminal VSC-{HVDC} Grids for Offshore Wind Farms},
  journal = {IEEE Transactions on Power Delivery},
  volume  = {26},
  number  = {4},
  pages   = {2476--2485},
  year    = {2011}
}

@article{Thams2017_DroopStructures_TPWRD,
  author  = {F. Thams and R. Eriksson and M. Molinas},
  title   = {Interaction of Droop Control Structures and Its Inherent Effect on the Power Transfer Limits in Multiterminal VSC-{HVDC}},
  journal = {IEEE Transactions on Power Delivery},
  volume  = {32},
  number  = {1},
  pages   = {182--192},
  year    = {2017}
}

@article{Rouzbehi2014_OPF_Droop_JESTPE,
  author  = {K. Rouzbehi and A. Miranian and A. Luna and P. Rodriguez},
  title   = {DC Voltage Control and Power Sharing in Multiterminal DC Grids Based on Optimal {DC} Power Flow and Voltage-Droop Strategy},
  journal = {IEEE Journal of Emerging and Selected Topics in Power Electronics},
  volume  = {2},
  number  = {4},
  pages   = {1171--1180},
  year    = {2014}
}

@ARTICLE{Mahmood_SST_2026,
  author={Mahmood, Hisham and Elizondo, Marcelo A.},
  journal={IEEE Transactions on Smart Grid}, 
  title={Grid‑Forming‑Capable Coordination of SST‑Interconnected Microgrids}, 
  year={2026},
  volume={},
  number={},
  pages={1-1},
  doi={10.1109/TSG.2026.3732228}}

@article{zhang_tpwrs_2010_psc,
  author   = {L. Zhang and L. Harnefors and H.-P. Nee},
  title    = {Power-Synchronization Control of Grid-Connected Voltage-Source Converters},
  journal  = {IEEE Transactions on Power Systems},
  volume   = {25},
  number   = {2},
  pages    = {809--820},
  year     = {2010},
  doi      = {10.1109/TPWRS.2009.2032231}
}

@article{Zhong2011_Synchronverter_TIE,
  author  = {Q.-C. Zhong and G. Weiss},
  title   = {Synchronverters: Inverters that Mimic Synchronous Generators},
  journal = {IEEE Transactions on Industrial Electronics},
  volume  = {58},
  number  = {4},
  pages   = {1259--1267},
  year    = {2011}
}

@inproceedings{dhople_allerton_2013_voc,
  author    = {S. V. Dhople and B. B. Johnson and A. O. Hamadeh},
  title     = {Virtual Oscillator Control for Voltage Source Inverters},
  booktitle = {Proc. 51st Annual Allerton Conference on Communication, Control, and Computing},
  pages     = {1359--1363},
  year      = {2013},
  doi       = {10.1109/Allerton.2013.6736685}
}

@article{Johnson2016_VOC_TPEL,
  author  = {B. B. Johnson and S. V. Dhople and A. S. Hamadeh and P. T. Krein},
  title   = {Synchronization of Islanded Inverter-Based Networks Using Virtual Oscillator Control},
  journal = {IEEE Transactions on Power Electronics},
  volume  = {31},
  number  = {4},
  pages   = {3607--3620},
  year    = {2016}
}

@article{Zhang2011_PSC_Weak_TPWRS,
  author  = {L. Zhang and L. Harnefors and H.-P. Nee},
  title   = {Interconnection of Very Weak AC Systems by VSC-HVDC Links Using Power-Synchronization Control},
  journal = {IEEE Transactions on Power Systems},
  volume  = {26},
  number  = {3},
  pages   = {344--355},
  year    = {2011}
}

@article{wu_tie_2019_psc_stability,
  author   = {H. Wu and X. Wang},
  title    = {Design-Oriented Transient Stability Analysis of Grid-Connected Converters With Power Synchronization Control},
  journal  = {IEEE Transactions on Industrial Electronics},
  volume   = {66},
  number   = {8},
  pages    = {6473--6482},
  year     = {2019},
  doi      = {10.1109/TIE.2018.2875669}
}

@techreport{cigre_tb604_2014,
  author      = {{CIGR{\'E} Working Group B4.57}},
  title       = {Guide for the Development of Models for {HVDC} Converters in a {HVDC} Grid},
  institution = {CIGR{\'E}},
  year        = {2014},
  number      = {Technical Brochure 604},
  url         = {https://www.e-cigre.org/publications/detail/604-guide-for-the-development-of-models-for-hvdc-converters-in-a-hvdc-grid.html},
  note        = {Accessed 2025-09-04}
}

@ARTICLE{Iraniparast_2025,
  author={Iraniparast, Amirhossein and Gro{\ss}, Dominic},
  journal={IEEE Transactions on Control of Network Systems},
  title={Networked Dynamics With Application to Frequency Stability of
  Grid-Forming Power-Limiting Droop Control},
  year={2025},
  volume={12},
  number={4},
  pages={2856--2867},
  month={Dec.},
  doi={10.1109/TCNS.2025.3591856}
}

\end{document}